\documentstyle[preprint,aps]{revtex}
\begin{document}
\preprint{\begin{tabular}{c}
\hbox to\textwidth{August 1997 \hfill BROWN-HET-1092}\\[-10pt]
\hbox to\textwidth{ \hfill hep-ph/9708409}\\[-10pt]
\end{tabular}}
\draft
\title{ The Minimal Supersymmetric Standard Model and Precision of 
$W$-Boson Mass and Top Quark Mass
}
\author{Kyungsik Kang and Sin Kyu Kang}
\address {\it Department of Physics, Brown University, 
 Providence, RI 02912, USA} 
%\date{\today}
\maketitle
\begin{abstract}
We argue that the present value and accuracy of $M_W$ and $m_t$ measurements
tend to favor the MSSM over the SM.
By speculating that a precision of the order 40 MeV and 3 GeV respectively
for $M_W$ and $m_t$ will be achieved at LEP2 and Tevatron,
we show that the prospect for the MSSM will be further enhanced as long as
the central values of $M_W$ and $m_t$ do not decrease below the present values.
In addition we discuss how this scenario can
constrain the Higgs boson mass and distinguish the Higgs boson of the MSSM
type from that of the SM.
\end{abstract}
%\pacs{PACS numbers: }

%\narrowtext
\newpage
Recent LEP measurements \cite{lep1,lep2}
have improved so precise that LEP's sensitivity
can even detect the passing of TGV train. 
The $W$-boson mass measured at LEP2, when combined with those at the
Tevatron, enables us to narrow $M_W$ within $\pm 0.08$ GeV \cite{lep2}, 
while the top 
quark mass has been measured also with a significantly smaller error
$\pm 5.5$ GeV \cite{top}.
These experimental advances should enable us to examine the effect of
the higher order radiative corrections more closely
and also  the existence of the Higgs boson for which we do not have direct 
evidence yet.
Even the apparent deviation of some of the electroweak parameters
such as $R_b, R_c$ and $A_{LR}$(SLD) from the standard model (SM) predictions
\cite{rb}, which persisted for several years and caused 
many theorists \cite{anom} to interpret
these anomalies as a possible signal of new physics beyond the SM, seems
to have been mostly washed away
from the very recent precision data from LEP and SLD, thus  making
the SM more appealing.
Although the measurements of most electroweak parameters may appear
to be consistent with the SM prediction,
we would like to point out in this Letter that the present value and
accuracy of $M_W$ 
and $m_t$ measurements \cite{lep2,top} tend to
support the MSSM rather than the SM.
Strictly speaking, the precision tests until now have resulted only a
consistent correlations among the relatively less accurate parameters
$M_W$ and $m_t$ and the unknown parameter $m_H$ within the framework of
the SM.
As we discussed in Ref. \cite{kk1}, because of the strong correlation
between $M_W$ and $m_t$ for a wide range of $m_H$, the future precision 
measurements
of $M_W$ amd $m_t$ beyond the current experimental accuracy
would provide a decisive and perhaps the only crucial test for or against
the SM and 
give a profound implication for the origin of the Higgs boson as well as
to the Higgs boson searches \cite{hunter}.
In this respect, the determinations of $M_W$ and  $m_t$
from $W^{+}W^{-}$ \cite{lep1,thres} and future $t\bar{t}$ threshold 
measurements \cite{thres2} 
could also play 
an important role on the future precision test of the SM and 
on the determination of the Higgs boson mass,
 because such measurements can severely reduce
the background and systematic uncertainties.
A precision of order $40$ MeV  for $M_W$ and 3 GeV for $m_t$ could perhaps be
achieved at the LEP2 as well as at the Tevatron \cite{lep3,lep4} soon.

In this Letter, we would like to discuss how such precise determinations
of $M_W$ and $m_t$ could serve to  test  the SM and constrain the
indirect bound of the Higgs boson mass.
In particular, we would like to point out that the current data tends
to favor already
the MSSM bounds over the SM type.
We will also examine how the future determinations of
$M_W$ and $m_t$ with further improved accuracy can provide a decisive clue 
on the evidence of a new physics 
beyond the SM, such as supersymmetry (SUSY).

As shown in Table I, the precision of $M_W$ and $m_t$ measurements has 
been steadily increased during the past years.
While the central value of $m_t$ is converging to around 175 GeV,
that of $M_W$ has been continuously increasing with improved accuracy.
As will be shown later, the increase of the central value of $M_W$ accompanied 
by the decrease of experimental uncertainty is potentially interesting
since it might at last lead to an evidence of new physics beyond the SM.
In Fig.1, we present the correlation between $M_W$ and $m_t$ in the SM 
(solid lines) and the MSSM (dashed lines).
Also shown in Fig.1 is the least-fit solutions ($-\diamond-$)
of the global fits to the SM for $m_H=100-1000$ GeV.
The prediction of $M_W$ versus $m_t$ is achieved with
$\alpha^{-1}(M_Z)=128.896\pm 0.09$ \cite{alpha} and 
$\alpha_s(M_Z)=0.118\pm 0.003 $ \cite{alphas}
from treating the relation of the radiative corrections self-consistently,
following the scheme described in Ref.\cite{kk1}.
The errors in $\alpha^{-1}(M_Z)$ and in $\alpha_s(M_Z)$ alone yield uncertainties on $M_W$ of order 20 MeV and 2 MeV, respectively.
In Fig.1, the band of SM prediction is obtained by taking the range
of $m_H$ to be $70~\mbox{GeV}\lesssim m_H \lesssim 1$ TeV, where
we use the current experimental lower limit of $m_H$ \cite{lep2}.
On the other hand, the lower bound on $m_H$ can theoretically be determined
from the vacuum stability condition within the context of the SM \cite{mh2}.
The lower bound on $m_H$ from the vacuum stability condition
depends on $m_t$ and new physics 
scale $\Lambda $ beyond which the SM is no longer valid.
This lower bound on $m_H$ increases with $m_t$ and decreases as $\Lambda$
is increased.
For the numerical values, we use the recent fit \cite{mh3}
\begin{equation}
m_H(\mbox{GeV}) > 133 + 1.92 (m_t(\mbox{GeV})-175)-4.28 \frac{\alpha_s(M_Z)
-0.12}{0.006}
\end{equation}
for $\Lambda=10^{19}$ GeV.
We note that the vacuum stability lower bound on $m_H$ becomes higher
than the current experimental lower limit for $m_t \gtrsim 145$ GeV.
Thus we take experimental bound in the
region  $m_t \lesssim 145$ GeV as the lower limit of $m_H$ 
and the vacuum stability lower bound for 
$m_t\gtrsim 145$ GeV in Fig.1.
%The upper limit of SM prediction corresponds to the experimental lower bound
%on $m_H(\simeq 70$ GeV) for $m_t\lesssim 145$ GeV and the vacuum stability
%lower bound on $m_H$ for $m_t\gtrsim 145$ GeV.
The upper line of the SM band corresponds to these lower bounds on $m_H$ in
two regions of $m_t$.
Thus, in the higher region of $m_t$, the upper limit of SM prediction for $M_W$
is somewhat lower than the one used by others in the literature 
\cite{lep4,graph}.
The lower line of the SM prediction corresponds to $m_H=1$ TeV as usual.
The MSSM bounds have been calculated by varying the SUSY parameter so that
they are consistent with current experimental results of the non-observation
of Higgs and SUSY particles at LEP2 \cite{lep4,graph,wsusy}.
In order to see the implication of the present measurements of $M_W$ and
$m_t$, we also present the current experimental results \cite{lep2,top}, 
$M_W=80.43\pm 0.08$ GeV and $m_t=175.6\pm 5.5 $ GeV in this figure.
The best global fit solutions give $M_W=80.331\pm0.024 (\mbox{due to}~m_H)
\pm0.020(\mbox{due to}~\alpha 's)$ GeV, somewhat lower than the current world
average of $M_W$.
Although the combined central point of the set $(M_W, m_t)$ tends to prefer 
the MSSM, 
both the SM and the MSSM  may actually be
consistent with the present accuracy of $M_W$ and $m_t$
in view of the uncertainties resulting from those in other input parameters
such as $\alpha^{-1}(M_Z), \alpha_s(M_Z),$ etc. as mentioned above.

From Fig.2, we see that the Higgs boson mass can be constrained by the present
experimental values of $M_W$ and $m_t$ by $m_H\lesssim 370$ GeV in general.
This is a remarkable improvement in the situation compared to the state of the
art in 1995 when we suggested to study $M_W$ vs. $m_H$ correlation for the first
time in Ref.\cite{kk1}.
The central values of $M_W$ and $m_t$ allow $m_H\simeq 67$ GeV  but with
an uncertainty $\Delta m_H$ of the order 130 GeV due to the experimental
uncertainties $\Delta M_W=80$ MeV and $\Delta m_t=5.5$ GeV.
This is consistent with the recent indirect determination of
$m_H=121^{+119}_{-68}$ GeV based on the global fit to the most recent
electroweak data 
\footnote{ Rerun of our ZFITTER program \cite{zft} 
following the scheme in Ref.\cite{kk1} has reproduced the same results.}
\cite{lep2}.

Now, let us speculate how the future measurement of $M_W$ and $m_t$ 
with improved accuracy can provide a decisive test for the SM:
let us assume that the central values of $M_W$ and $m_t$
would not decrease from the current values and that the precision of $M_W$ 
and $m_t$ would be improved to the order of
40 MeV and 3 GeV respectively, which will be achieved
at the LEP 2 and the Tevatron in the foreseeable future \cite{lep3}.
As one can see from Fig.1, this scenario seems to disfavor the SM
compared to the MSSM.
If the increasing trend of the central value of $M_W$ is to continue
above the present one,
the precision measurement of $M_W$ and $m_t$ would provide more distinctive 
evidence for new physics beyond the SM, in particular, in favor of the MSSM.
Thus, as long as the central value of $M_W$ is determined to increase
above the present value, the future precise measurements of $M_W$ and $m_t$ 
could serve a useful window to witness the cracks in the SM and
to look for the evidence of new physics.
On the other hand, as can be seen from Fig.3, the speculation to have
$M_W=80.43\pm0.04$ GeV and $m_t=175.6\pm 3$ GeV leads to
$m_H\lesssim 180$ GeV, in the framework of  the SM, with an uncertainty 
$\Delta m_H$ of the order 50 GeV due to the
uncertainties $\Delta M_W=40$ MeV and $\Delta m_t=3$ GeV.
However, as shown in Fig.1, this scenario seems to prefer the MSSM even better.
As is well known, in the MSSM, the intrinsic upper bound on the lightest
Higgs boson mass can be obtained by assuming that SUSY effects are
decoupled and the lightest Higgs boson is equivalent to the
SM Higgs boson below SUSY breaking scale \cite{mh3,mhsusy2}.
In the considered range of $m_t$, this upper bound turns out to be
$m_H \lesssim 130$ GeV \cite{mh3,mhsusy2} which is to be 
discriminated from the lower bound on 
the SM Higgs boson mass ($\gtrsim 133 $ GeV).
Thus, if the Higgs boson will be discovered in the range  
$m_H\lesssim 130$ GeV and
$M_W$ and $m_t$ will be measured with the above speculated values
and accuracy,
this  will constitute a definite signal in favor of the MSSM.
%
%Thus, if the Higgs boson will be discovered at $m_H \lesssim 180$ GeV and the
%precision $\Delta M_W \lesssim 40$ MeV and $\Delta m_t\lesssim 3$ GeV will be achieved without the decrease of the central value of $M_W$,
%the discovery of the Higgs scalar can imply the existence of SUSY.

We note that the upper limit of the SM prediction shown in Fig.1 corresponds to
the central values of $\alpha^{-1}(M_Z)$ and $\alpha_s(M_Z)$.
Thus, the upper limit can be increased as much as the lower limit
of the MSSM prediction,
if the errors in $\alpha^{-1}(M_Z)$ and $\alpha_s(M_Z)$ are included.
However, if the central value of $M_W$ is to increase by
20 MeV from the present value, the MSSM will again be preferred.

In conclusion, we have pointed out that the current value and
accuracy of $M_W$ and $m_t$ measurements tend to favor the MSSM
rather than the SM, although most recent elecroweak data
may appear to be consistent with the SM prediction. 
We have shown that our speculated values of $M_W$ and $m_t$ with
the precision of order
40  MeV and 3 GeV respectively, which could perhaps be achieved
at the LEP2 as well as the Tevatron in the foreseeable future,
could prefer the MSSM even better and
provide a decisive clue on the evidence of the MSSM
when the Higgs boson would be discovered in the range $m_H\lesssim 130$ GeV.
%In this case, the precision of $M_W$ and $m_t$ at least of order 20 MeV and 1 GeV would 
%be required in order to see any evidence of new physics.
%
%
%
%
\acknowledgements
One of us (S.K.K.) would like to thank the Korea Science and Engineering
Foundation for financial support and also the members of the High Energy
Theory Group for the warm hospitality extended to him at Brown University.
This work was supported in part by the U.S. DOE Contract No. DE-FG02-91ER 40688-Task A.

\begin{table}
\caption{The evolution of $M_W$ and $m_t$ measurements.}
\label{table I}
\begin{tabular}{ccc}
& $M_W$ (GeV)& $m_t$ (GeV) \\
\tableline
%& & \\
1997 (Summer) \cite{lep2} & $ 80.43\pm 0.08$ & $175.6\pm 5.5$ \\
1997 (January) \cite{lep1} & $ 80.37\pm 0.08$ & $175.6\pm 5.5$ \\
1996 (Summer) \cite{lep5} & $ 80.356\pm 0.125$ & $175\pm 6$ \\
1995 (Summer) \cite{lep6} & $ 80.26\pm 0.16$ & $180\pm 13$ \\
1994 (Summer) \cite{lep7} & $ 80.23\pm 0.18$ & $174\pm 10^{+18}_{-12}$(CDF) \\
%\tableline \tableline
\end{tabular}
\end{table}
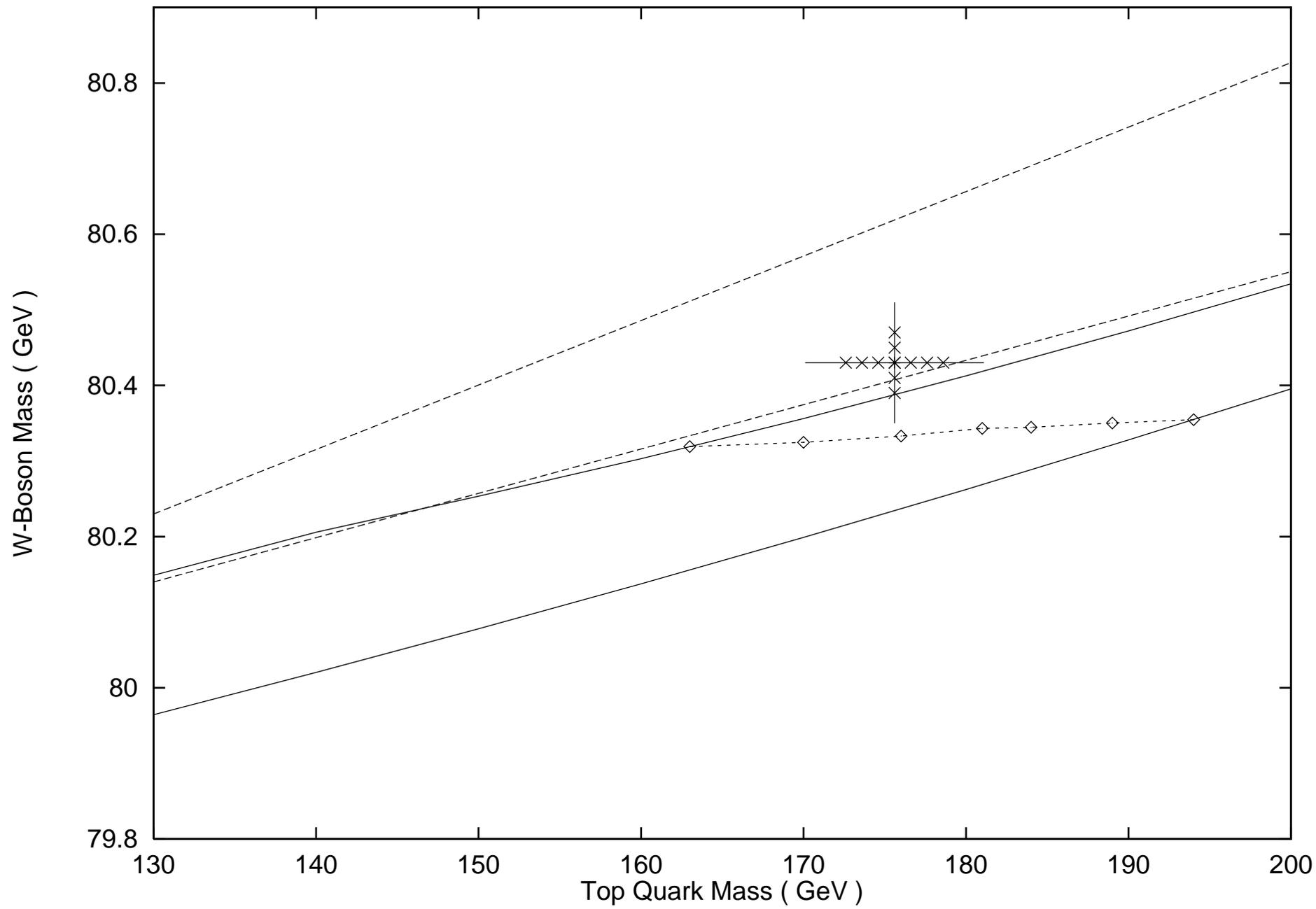
\begin{figure}
\caption{$M_W$ versus $m_t$ in the SM (solid lines) and the MSSM (dashed lines).
The current experimental results, $M_W=80.43 \pm 0.08$ GeV and 
$m_t=175.6\pm 5.5$ GeV are presented and the speculation to improve
the errors by half so that we have
$M_W=80.43\pm 0.04$ GeV and $m_t=175.6\pm 3.0$ GeV are also indicated
by the cross points (xxx).
The cases of the minimal $\chi^2$-fit from the global fits to the recent LEP data are shown
by $\diamond$.}
\label{1}
\end{figure}
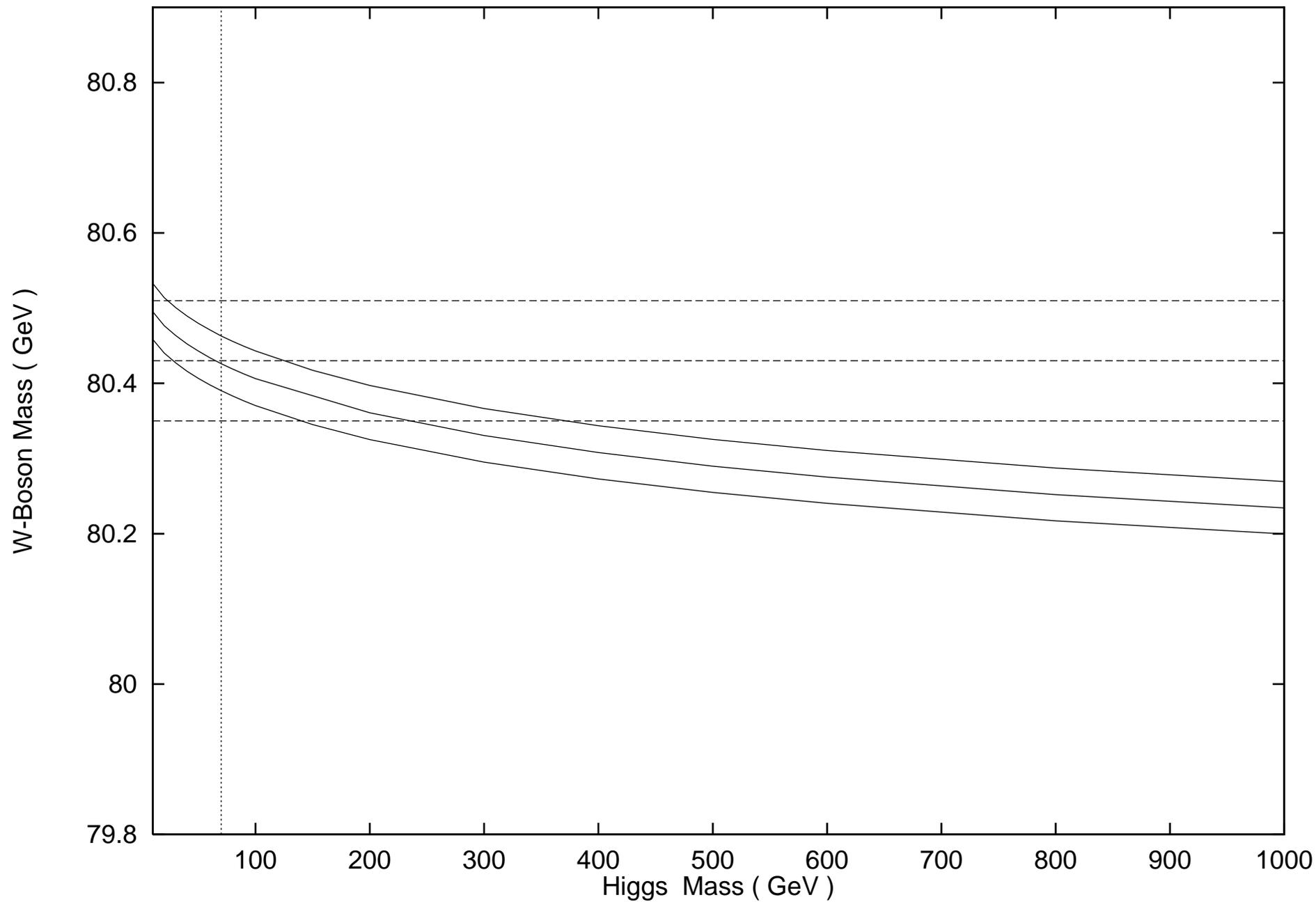
\begin{figure}
\caption{$M_W$ versus $m_H$ for $ m_t=175.6\pm5.5$ GeV.
The current experimental measurement of $M_W$ is shown by dashed lines.
The vertical dotted line corresponds to the experimental lower limit
of the Higgs boson mass ($m_H=70$ GeV).}
\label{2}
\end{figure}
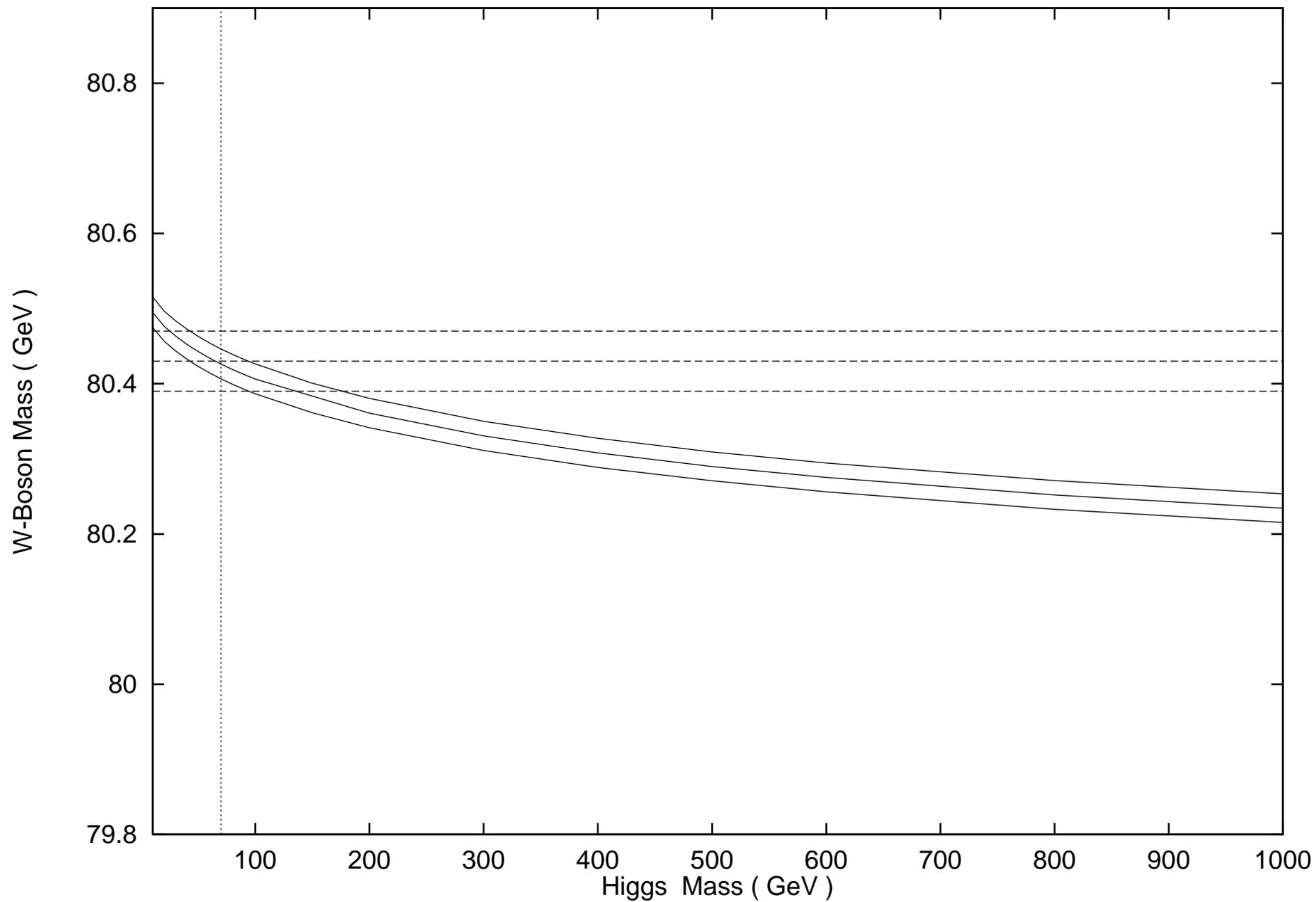
\begin{figure}
\caption{The same as Fig.2 but for $ m_t=175.6\pm3.0$ GeV and
$M_W=80.43\pm 0.04$ GeV.}
\label{3}
\end{figure}
\end{document}